\documentclass[preprint,showpacs,amsmath,amssymb]{revtex4}

\begin{document}
\def\bea{\begin{eqnarray}}
\def\eea{\end{eqnarray}}
\def\nn{\nonumber}
\renewcommand\epsilon{\varepsilon}
\def\beq{\begin{equation}}
\def\eeq{\end{equation}}
\def\lla{\left\langle}
\def\rra{\right\rangle}
\def\za{\alpha}
\def\zb{\beta}
\def\lsim{\mathrel{\raise.3ex\hbox{$<$\kern-.75em\lower1ex\hbox{$\sim$}}} }
\def\gsim{\mathrel{\raise.3ex\hbox{$>$\kern-.75em\lower1ex\hbox{$\sim$}}} }
\newcommand{\Rbs}{\mbox{${{\scriptstyle \not}{\scriptscriptstyle R}}$}}

\draft

\title{Implications of SNO and BOREXINO results \\
on Neutrino Oscillations and Majorana Magnetic Moments
}

\author{ S.~ K.~ Kang$^{a}$\thanks{ E-mail : skkang@phya.snu.ac.kr}~ and~
C.~ S.~ Kim$^{b}$ \thanks{E-mail : cskim@yonsei.ac.kr}}
\date{\today}
\address{\small \it $^{a}$School of Physics, Seoul National
University,
       Seoul 151-734,Korea \\
       $^{b}$Department of Physics, Yonsei University, Seoul 120-749, Korea}

\begin{abstract}

\noindent Using the recent measurement of SNO salt phase experiment, we investigate
how much the solar neutrino flux deficit observed at SNO could be due to
$\nu_e$ transition into antineutrino.
Our analysis leads to
rather optimistic conclusion that the SNO salt phase data may indicate
the existence of Majorana magnetic moment.
The prospect for the future  BOREXINO experiment is also discussed.

\end{abstract}
\pacs{ 14.60.Pq, 14.60.St, 13.40.Em} \maketitle
\thispagestyle{empty} \narrowtext
In addition to the solar neutrino experiment at Super-Kamiokande (SK)
\cite{SK2002},  the recent neutrino experiments at Sudbury Neutrino
Observatory (SNO) \cite{SNOCC,SNONC,SNOS} and KamLAND
\cite{kamex} indicate that the long-standing solar neutrino problem,
discrepancy between the prediction of the neutrino flux
based on the standard solar model (SSM) \cite{ssm} and that
measured by experiments, can be resolved in terms of neutrino oscillations.
Both the experiments, SNO and SK,  probe the high energy tail of the solar
neutrino spectrum, which is dominated by the $^8B$ neutrino flux.
The water Cerenkov experiments from Super-Kamiokande
(SK) \cite{SK2002} has observed the emitted electron from elastic
scattering ($ES$) $ \nu_{x} + e \rightarrow \nu_{x} + e,
\label{one} $ ($\nu_x=\nu_e,\nu_{\mu},\nu_{\tau}$), while SNO has
measured the neutrino flux  through the charged current ($CC$)
process $\nu_e + d \rightarrow p + p + e \label{two},$ the neutral
current ($NC$) process $ \nu + d \rightarrow \nu + p + n,
\label{three} $ and $ES$ process given in the above.
Very recently, SNO has measured the total active $^8B$ solar neutrino flux
with dissolved NaCl in the heavy water to enhance the sensitivity
and signature for $NC$ interactions \cite{SNOS}.
The results of the solar neutrino flux measured at SK and SNO are
given in Table 1. Note that the  SNO salt data I in Table 1 presents
solar neutrino fluxes detected through $CC, ES$ and $NC$ without
the constraint of an undistorted $^8B$ energy spectrum, while the
SNO salt data II presents solar neutrino fluxes by adding the
constraint.
Based on a global analysis in the framework of two-active neutrino
oscillations of all solar neutrino data and KamLAND result,
the large mixing angle
(LMA) solution is favored and oscillations into a pure sterile
state are excluded at high confidence level \cite{lma}.
It also appears that all non-oscillation solutions of the solar neutrino
problem are strongly disfavored \cite{kamth,decay}.

The spin flavor precession (SFP) solution of the solar neutrino
problem \cite{SFP}, motivated by the possible existence of nonzero
magnetic moments of neutrinos, has attracted much attention before
the KamLAND experiment. Although the KamLAND result excludes a
{\it pure} SFP solution to the solar neutrino problem under the
CPT invariance, a fraction of the flux suppression of solar
neutrino may still be attributed to SFP \cite{comb1}. In this
respect, we believe that the detailed investigation on how much
the flux suppression of solar neutrino can be attributed to SFP
will lead us to make considerable progress in understanding the
solar neutrino anomaly as well as the inner structure of the Sun.
In addition, the observation of solar active antineutrino flux
must be a signature for the existence of  Majorana neutrinos and
working of SFP inside the Sun \cite{SFP,dev}. Recently, we
have investigated a possibility to resolve the solar neutrino anomaly
observed from the solar neutrino experiments in terms of the
combination of the neutrino oscillations and the neutrino
spin-flavor conversions \cite{skk}.
To achieve our goal, we have proposed a simple and
model-independent method to extract information on $\nu_e$
transition into antineutrinos via SFP from the measurements of
$^8B$ neutrino flux at SNO and SK, and showed how much the solar
neutrino flux deficit observed at SNO and SK could be due to
$\nu_e$ transition into antineutrino. As has been seen, in
particular, our determination of the mixing between non-electron
active neutrino and antineutrino is not affected by the existence
of transition into a sterile state \cite{skk}.

\begin{table}
\begin{tabular}{|ccc|}\hline
Experiment& (interaction : flux $\Phi_{\rm exp}^{int}$)  &  \\
\hline
 SK & $(ES:2.35\pm 0.08)$  & \\
 \hline
Old SNO & $(ES : 2.39\pm 0.27 )$  &\\
        &$(CC :1.76\pm 0.11) $
         &$(NC: 5.09\pm 0.62) $ \\
\hline  SNO salt I & $(ES : 2.21\pm 0.30 )$ & \\
        &$(CC :1.59\pm 0.11) $
         &$(NC: 5.21\pm 0.47) $ \\
\hline  SNO salt II & $(ES : 2.13\pm 0.32 )$  & \\
        &$(CC :1.70\pm 0.11) $
         &$(NC: 4.90\pm 0.37) $ \\ \hline
\end{tabular}
 \caption
 {Solar neutrino flux measured at SK
 and SNO
 in the unit of $10^6
\mbox{cm}^{-2}\mbox{s}^{-1}$ .
 }
\end{table}

In this letter, we shall update the analysis based on the recent measurement
of SNO salt phase experiment and investigate how large
the transition of solar $\nu_e$ into non-electron antineutrinos could be
responsible for the deficit of solar neutrino flux.
As will be shown, our analysis leads to
rather optimistic conclusion that the SNO salt phase data may indicate
the existence of Majorana magnetic moment.

Let us begin by considering how the experimental measurement of
the solar neutrino flux can be presented in terms of the solar
neutrino survival probability.
The excess of $NC$ and $ES$ can be caused not only by the active neutrinos
but also by the active antineutrinos.
The antineutrinos in question
are mostly of the muon or tau types because of no observation of
$\bar\nu_e$ \cite{antiel}.
 Both $\nu_{\mu,\tau}$ and
$\bar\nu_{\mu,\tau}$ scatter on electrons and deuterium nuclei
through their $NC$ interactions, with different cross sections.
Assuming the SSM neutrino fluxes, $\Phi_{\rm SSM}=5.05^{+1.01}_{-0.81}
\times 10^6 \mbox{cm}^{-2} \mbox{s}^{-1}$, and
the transition of $\nu_e$ into a mixture of  active (anti-)flavor
$\nu_{a(\bar{a})}$
and sterile $\nu_s$
that
participate in the solar neutrino oscillations, one can write the
SNO $ES$, $CC$ and $NC$ scattering rates relative
to the SSM predictions in terms of the survival probability
\cite{barger,snofit}:
\begin{eqnarray}
R^{ES}_{\rm SNO} &\equiv& \Phi^{ES}_{\rm SNO} /\Phi_{\rm SSM}
               = f_B \left[P_{ee} +
r\sin^2\alpha \sin^2 \psi (1-P_{ee}) \right. \nonumber \\
 & & ~~~~~~~~~~~~~~~\left. +~~~ \bar{r}\sin^2 \alpha \cos^2
\psi (1-P_{ee})\right], \label{rel2} \\
R^{CC}_{\rm SNO} &\equiv & \Phi^{CC}_{\rm SNO} /\Phi_{\rm SSM}
               = f_B P_{ee},
\label{seven} \\
R^{NC}_{\rm SNO} &\equiv & \Phi^{NC}_{\rm SNO} /\Phi_{\rm SSM}
                = f_B [P_{ee} +\sin^2\alpha (1-P_{ee}) ],
\label{eight}
\end{eqnarray}
where
$r \equiv \sigma^{NC}_{\nu_a}/\sigma^{CC+NC}_{\nu_e} \simeq 0.154$
and
$\bar{r} \equiv \sigma^{NC}_{\bar{\nu}_a}/\sigma^{CC+NC}_{\nu_e} \simeq
0.114$
for a threshold energy of 5 MeV \cite{bahcall}, and $P_{ee}$ is
the $\nu_e$ survival probability. Here $\sin^2\alpha$ indicates
the fraction of
 $\nu_e$ oscillation to active flavor $\nu_a$, whereas
$\psi$ is a mixing angle that describes the linear combination of
the probabilities of $\nu_e$ conversion into $\nu_a$ and
$\bar{\nu}_a$ .
Since there is a large uncertainty in the predicted
normalization of $\Phi_{\rm SSM}$,
arising from the uncertainty in the $^7Be + p \rightarrow \ {^8B}
+ \gamma $ cross-section, we have introduced a constant parameter
$f_B$ to denote the normalization of the $^8B$ neutrino flux
relative to the SSM prediction.
We assume a common survival probability for all the three
measurements.
Using the measured values of the rates $R$, we can estimate the
allowed regions of the quantities $f_B, P_{ee}$ and $\alpha$. In
particular, the fraction of $\nu_e$ oscillation to $\nu_a$ is
described by the relation \cite{barger,snofit},
\begin{equation}
\sin^2\alpha = \frac{R_{\rm SNO}^{NC}-R_{\rm SNO}^{CC}}
{f_B-R^{CC}_{\rm SNO}}. \label{rel1}
\end{equation}
Imposing the SSM constraint $f_B=1\pm 0.18$ \cite{barger} and the
experimental results for the ratios $R^\prime s$, we obtain
\begin{equation} \sin^2\alpha = 1.05\pm 0.32.
\label{rel11}
\end{equation}
We see that the evidence for transitions to active neutrinos is at
the $3.3\sigma$ C.L., but large sterile fractions are still
allowed.

{}From Eqs. (\ref{rel2},\ref{seven},\ref{eight}), we see that the
mixing angle $\psi$ is related with the measured neutrino fluxes
as follows:
\begin{eqnarray}
r\sin^2\psi + \bar{r}\cos^2\psi =
 \frac{R^{ES}_{\rm
SNO}-R^{CC}_{\rm SNO}}{R^{NC}_{\rm SNO}-R^{CC}_{\rm SNO}} ,
\label{rel3}
\end{eqnarray}
where we have assumed that $\sin^2\alpha$ is non-zero. The
expression (\ref{rel3}) shows that  the determination of the
mixing angle $\psi$ is independent of nontrivial $\sin^2\alpha$,
and the precise measurements
of $R^{ES}_{\rm SNO},R^{NC}_{\rm SNO},R^{CC}_{\rm SNO}$ as
well as the values of $r$ and $\bar{r}$ make it possible to see
how much the solar neutrino flux deficit can be caused by SFP. We
note that any deviation of the value of $\sin^2\psi$ from one
implies the evidence for the existence of $\nu_{e}$ transition
into non-sterile antineutrinos, and if there is no transition of
solar neutrino due to the magnetic field inside the sun, the
left-hand side of Eq. (\ref{rel3}) should be identical to the
parameter $r$. To obtain the values for the right-hand side of Eq.
(\ref{rel3}), we  consider two combinations of the
experimental results measured through $CC,ES$ and $NC$
interactions :
\begin{eqnarray}
{\rm (a)~~SNO~salt~data~phase~I} &:&
(\Phi_{\rm SNO}^{CC},~\Phi_{\rm SNO}^{ES},~\Phi_{\rm SNO}^{NC}), \nonumber\\
{\rm (b)~~SNO~salt~data~phase~II} &:&
(\Phi_{\rm SNO}^{CC},~\Phi_{\rm SNO}^{ES},~\Phi_{\rm SNO}^{NC}), \nonumber
\end{eqnarray}
and then
the results are given as follows:
\begin{eqnarray}
{\rm Eq.~(6)}  \Rightarrow \left\{\begin{array}{c}
 (\mbox{a})~~~0.171 \pm 0.089~,  \\
  (\mbox{b})~~~0.134 \pm 0.105~. \end{array} \right.
\label{num}
\end{eqnarray}
Since $\bar{r} \leq r\sin^2\psi + \bar{r}\cos^2\psi \leq r$, we
notice that the left-hand side of Eq. (\ref{rel3}) prefers to
lower sides of Eq. (\ref{num}), and
leads to
\begin{eqnarray}
\sin^2\psi = \left\{ \begin{array}{c} (\mbox{a})~~~1.43 \pm 0.33~, \\
  (\mbox{b})~~~0.51 \pm 0.39~.\end{array} \right.
\label{numa}
\end{eqnarray}
{}From Eq. (\ref{numa}), we see that  both pure active neutrino
oscillation and neutrino oscillation$+$SFP are allowed  for the SNO salt
data I (combination (a)) within $2 \sigma$
level, whereas the SNO salt data II ( combination (b)) shows that the
existence of solar $\nu_e$ transition into $\bar{\nu}_a$ is at $1.3 \sigma$
although the pure active neutrino oscillation is allowed within
$2\sigma$. Therefore, only new SNO data constrained by undistorted
$^8B$ neutrino spectrum implies an evidence for the existence of
the spin-flavor transition due to Majorana neutrino magnetic
moment in the solar neutrino fluxes. We note that the main reason
for $\sin^2 \psi > 1.0$ in the case of (a) is due to large
deviation of $CC$ flux from $ES$ one.
Thus, the precise determination of the central values of each flux as
well as reduction of the uncertainties will lead us to
precisely probe the existence of
the solar neutrino transition into antineutrino in the above way.
In this analysis, we have taken into account only SNO salt phase results.
If we replace the SNO $ES$ rates with the SK $ES$ one as done in
\cite{skk}, the value of $\sin^2\psi$ comes out to be very large because
the flux of SK $ES$ is rather larger than those of SNO salt $ES$
as shown in Table I.

Let us briefly discuss the prospect for future experiment, BOREXINO,
which will detect  the medium energy $^{7}Be$,CNO and pep
solar neutrinos through $ES$ interaction \cite{bor}.
Assuming that the observed flux deficit of solar neutrinos
is due to the combination of neutrino oscillations and SFP transitions,
we can predict $R_{\rm BOR}^{ES}= \Phi_{\rm BOR}^{ES}/\Phi_{\rm SSM} $.
In order to do that, we first determine the survival probability
of the medium energy neutrinos by comparing Homestake event rate \cite{cl} with
the  SNO $CC$ result. Since the fractional contributions of the high energy
$^8B$ and the medium energy neutrinos to the $^{37}{\rm Cl}$ signals are
$76.4\%$ and $23.6\%$,respectively, the measured rate divided by the SSM
prediction for the Homestake experiment $R_{\rm Cl}$ with oscillations is
given by
\begin{equation}
R_{\rm Cl} = 0.764~ f_{B} P^{B}_{ee} + 0.236~ P^{M}_{ee}
\end{equation}
where $P^{B}_{ee}$ is the survival probability for $^8B$ neutrinos, whereas
$P^{M}_{ee}$ is that for the medium energy neutrinos.
Since $f_{B} P^{B}_{ee}$ is equivalent to $R^{CC}_{\rm SNO}$, we can obtain
the numerical value of $P^M_{ee}$ by using the experimental results for
$R^{CC}_{\rm SNO}$ and $R_{\rm Cl}$:
\begin{eqnarray}
 P^M_{ee}=\left\{ \begin{array}{c} (\mbox{a})~~~0.409 \pm 0.105~, \\
  (\mbox{b})~~~0.338 \pm 0.105~.\end{array} \right.
\end{eqnarray}
Similar to the SNO measured rates,
by allowing both neutrino oscillation and SFP transitions,
the BOREXINO $ES$ rate relative to the
SSM predictions in terms of the survival probability is
presented as
\begin{equation}
R_{\rm BOR}^{ES} = P^{M}_{ee}+\sin^2\alpha
(r\sin^2\psi +\bar{r}\cos^2\psi ) (1-P^M_{ee})~,
\end{equation}
where $r\simeq0.213$ and $\bar{r}\simeq 0.181$ for $^7Be$ neutrinos.
Using the above results Eqs.(\ref{rel11},\ref{numa}), we can obtain
\begin{eqnarray}
R_{\rm BOR}^{ES} =
 \left\{ \begin{array}{c} (\mbox{a})~~~0.549 \pm 0.117~, \\
  (\mbox{b})~~~0.475 \pm 0.116~.\end{array} \right.
\label{numa1}
\end{eqnarray}
It can be interesting to compare the above with the predictions for
pure neutrino oscillation cases ($\sin^2\psi=1$) which are given by
\begin{eqnarray}
R_{\rm BOR}^{ES} =
 \left\{ \begin{array}{c} (\mbox{a})~~~0.541 \pm 0.116~, \\
  (\mbox{b})~~~0.486 \pm 0.117~.\end{array} \right.
\label{numa2}
\end{eqnarray}

We note that the main uncertainties in (\ref{numa1},\ref{numa2}) are
due to the uncertainty in $P^M_{ee}$.
{}From the above results, we see that it might be difficult to
discriminate between pure oscillation solution and oscillation $+$ SFP
solution unless the future BOREXINO experiment measures $R^{ES}_{\rm BOR}$
with the uncertainty  $\delta R^{ES}_{\rm BOR}\sim 2-3\%$.
If the future SNO experiment could reduce the errors in the flux measurements
to about $50\%$, then the
uncertainty on $\sin^2\psi$ becomes $\delta \sin^2\psi \simeq (\mbox{a}) 0.17
~(\mbox{b}) 0.19$
and if the errors in
$P^M_{ee}$ could be reduced to $30\%$,
the uncertainties in the prediction for $R^{ES}_{\rm BOR}$ becomes
about $\delta R^{ES}_{\rm BOR} \simeq 0.05$ which is still a little large
to see whether there exists an evidence for the existence of spin-flavor
transition from BOREXINO experiment.
However, since oscillation $+$ SFP solution prefers lower value of
$R^{ES}_{\rm BOR}$, if the future BOREXINO will measure $R^{ES}_{\rm BOR}\leq
0.37$, it might be an indirect evidence for the existence of Majorana neutrinos
and working SFP mechanism in the Sun.
In addition, we hope that the future BOREXINO
experiment would make us to decide which case of SNO data set between (a) and
(b) is more relevant.

In summary, we have examined in a simple and model-independent way
how much the $\nu_e$ transition into antineutrinos could be in the
solar neutrino flux. The SNO salt data constrained by an
undistorted $^8B$ energy spectrum indicates the existence of
Majorana magnetic moment and working SFP mechanism within about $
1 \sigma $ level, while the SNO salt data without that constraint
allows both pure active neutrino oscillation and the effect of SFP
within $2\sigma $ level. The prospect for the future  BOREXINO
experiment has been discussed.

S.K.K is supported  by BK21 program of the Ministry of
Education in Korea. The work of C.S.K. was supported by Grant
No. R02-2003-000-10050-0 from BRP of the KOSEF.

\end{document}